%Paper: hep-th/9407049
%From: theory@qmchep.DNET.NASA.GOV (QMW THEORY GROUP)
%Date: Fri, 8 Jul 94 10:22:11 -0400

\input phyzzx
\voffset = -0.4in
\footline={\ifnum\pageno=1 \nulline \else\newfootline \fi}
\def\nulline{{\hfill}}
\def\newfootline{\advance\pageno by -1\hss\tenrm\folio\hss}
\rightline {May 1994}
\rightline {QMW--TH--94/18}
\title {MODULAR SYMMETRIES, THRESHOLD CORRECTIONS \break
AND  MODULI \  FOR $Z_2 \times Z_2 $ ORBIFOLDS.\break}
\author{D. Bailin$^{a *}$, \ A. Love$^{b}$,  \ W.A.
Sabra$^{b **}$\ and \ S. Thomas$^{c ***}$}
\address {$^{a}$School of Mathematical and Physical
Sciences,\break
University of Sussex, \break Brighton U.K.}
\address {$^{b}$Department of Physics,\break
Royal Holloway and Bedford New College,\break
University of London,\break
Egham, Surrey, U.K.}
\address {$^{c}$
Department of Physics,\break
Queen Mary and Westfield College,\break
University of London,\break
Mile End Road, London,  U.K.}
\footnote*{e-mail address: D.Bailin@SUSSEX.AC.UK.}
\footnote{**}{e-mail address: UHAP12@VAX.RHBNC.AC.UK.}
\footnote{***}{e-mail address: S.THOMAS@QMW.AC.UK.}
\abstract {${\bf Z}_2\times {\bf Z}_2$ Coxeter orbifolds are constructed with
the property that some twisted sectors have fixed planes for which the
six-torus can not be decomposed into a direct sum ${\bf T}^2\bigoplus{\bf T}
^4 $ with the fixed plane lying in ${\bf T}^2$.
The string loop threshold corrections to the gauge coupling constants are
derived, and display symmetry groups for the $T$ and $U$ moduli that are
subgroups of the full modular group $PSL(2,Z)$. The effective potential
for duality invariant gaugino condensate in the presence of hidden sector
matter is constructed and minimized for the values of the moduli. The effect
of Wilson lines on the modular symmetries is also studied.}
\endpage
%%%%%%%%%%%%%%%%%%%%%%%%%%%%%%%%%%%%%%%%%%%%%%%%%%%%%%%%%%%%%%%%%%%%%%%%%%%%
%%%%%%%%%%%%%%%%     REFS    %%%%%%%%%%%%%%%%%%%%%%%%%%%%%%%%%%%%%%%%%%%%%%%
%%%%%%%%%%%%%%%%%%%%%%%%%%%%%%%%%%%%%%%%%%%%%%%%%%%%%%%%%%%%%%%%%%%%%%%%%%%%
\REF\one{L. Dixon, J. A. Harvey, C. Vafa and E. Witten, Nucl. Phys.
B261 (1985) 678; B274 (1986) 285.}
\REF\two{ A. Font, L. E. Ibanez,
F. Quevedo and A. Sierra, Nucl. Phys. B331 (1991) 421.}
\REF\three {V. S. Kaplunovsky, Nucl. Phys. B307 (1988) 145}
\REF\four {L. J.
Dixon, V. S. Kaplunovsky and J. Louis,  Nucl. Phys. B355 (1991) 649.}
\REF\five{L. E. Ibanez, D. L\"{u}st and G. G. Ross, Phys. Lett. B272
(1991)
25.}
\REF\six{L. E. Ibanez and D. L\"{u}st , Nucl. Phys. B382 (1992) 305.}
\REF\seven{J. P. Derendinger, S. Ferrara, C. Kounas and
F. Zwirner, Nucl. Phys. B372 (1992) 145, Phys. Lett. B271 (1991)
307.}
\REF\eight{ D. Bailin and A. Love, Phys. Lett. B278 (1992) 125;
Phys. Lett. B292 (1992) 315.}
\REF\nine{  A. Font, L. E. Ibanez, D. Lust and F. Quevedo,
 {\it Phys. Lett. } {\bf B245} (1990) 401;
S. Ferrara, N. Magnoli, T. R. Taylor and G. Veneziano,
{\it Phys.Lett.} {\bf B245} (1990) 409;
M. Cvetic, A. Font, L. E. Ibanez, D. Lust and F. Quevedo,
{\it Nucl. Phys. } {\bf B361 } (1991) 194. }
\REF\ten{L. Dixon, talk presented at the A. P. S.  D. P. F. Meeting at
Houston
(1990); V. Kaplunovsky, talk presented at "Strings 90" workshop at College
Station (1990); L. Dixon, V. Kaplunovsky, J. Louis and M. Peskin,
unpublished.}
\REF\eleven{ J. A. Casas, Z. Lalak, C. Munoz and G. G. Ross, {\it Nucl.
Phys.}
{\bf  B347} (1990) 243.}
\REF\twelve{H. P. Nilles and M. Olechowsky, {\it Phys. Lett. } {\bf B248}
(1990) 268; P. Binetruy and M. K. Gaillard, {\it Phys. Lett. } {\bf B253}
(1991) 119.}
\REF\thirteen{ J. Louis, SLAC-PUB-5645 (1991).}
\REF\fourteen{D. Lust and T. R. Taylor, {\it Phys. Lett. } {\bf B253}
(1991) 335.}
\REF\fifteen{D. Lust and C. Munoz, {\it Phys. Lett. }{\bf B279}
(1992) 272;
B. de Carlos, J. A. Casas and C. Munoz,
{\it Nucl. Phys. } {\bf B339 } (1993) 623.}
\REF\sixteen{D. Bailin, A. Love, W.A.
Sabra and S. Thomas, preprint QMW--TH--94/11, SUSX--TH--94/11.}

\REF\seventeen{ P. Mayr and S. Stieberger, {\it Nucl. Phys}. {\bf B407}
(1993)
725.}
\REF\eightteen{D. Bailin, A. Love, W. A. Sabra and S. Thomas,
{\it Phys. Lett.} {\bf  B320} (1994) 21.}
\REF\nineteen{D. Bailin, A. Love, W. A. Sabra and S. Thomas, {\it Mod.
Phys.
Lett}. {\bf A 9} (1994) 67.}
\REF\twenty{M. Spalinski,  {\it Phys.  Lett}. {\bf B275} (1992) 47;
J. Erler, D. Jungnickel and H. P. Nilles, {\it Phys. Lett}. {\bf B276}
(1992)
303.}
\REF\twentyone{J. Erler and M. Spalinski, preprint, MPI-PH-92-61,
TUM-TH-147-92}
\REF\twentytwo{ D. Bailin, A. Love, W. A. Sabra and S. Thomas, QMW-TH-93-31,
SUSX-TH-93-17, to appear in {\it Mod. Phys. Lett.} {\bf A}.}
\REF\twentythree{ A. Love, W. A. Sabra and S. Thomas,
QMW-TH-94-05.}
\REF\twentyfour{Y. Katsuki, Y. Kawamura, T. Kobayashi, N. Ohtsubo, Y. Ono
and K. Tanioka, {\it Nucl. Phys}. {\bf B341} (1990) 611.}

The spectrum of states for an orbifold [\one, \two] is invariant under
discrete transformations for the $T$ and $U$ moduli together with the
winding numbers and momenta on the orbifold. These modular symmetries also
appear as symmetries of the string loop threshold corrections which are
important for the unification of gauge coupling constants [\three-\eight].
Moreover, the form of the threshold corrections, which is dictated to a
considerable extent by the modular symmetries, influences the form of the
non-perturbative superpotential due to gaugino condensation in the hidden
sector, and so the effective potential that determines the values of the $T$
and $U$ moduli [\nine-\fifteen].

In the absence of Wilson lines, provided all twisted sector fixed planes are
such that the six-torus can be decomposed into a direct sum ${\bf T}
^2\bigoplus {\bf T}^4 $ with the fixed plane lying in ${\bf T}^2$,
the group of modular symmetries of the threshold corrections is a product of
$PSL(2,Z)$ factors one for each of the $T$ and $U$ moduli associated with
the fixed planes. We shall refer to such orbifolds as decomposable orbifolds.
However, for non-decomposable orbifolds, the group of modular symmetries is
in general a product of congruence subgroups [\seventeen-\nineteen] of
$PSL(2,Z)$. Wilson lines can also break the $PSL(2,Z)$ modular symmetries
[\twenty-\twentythree].

The modular symmetry groups for string loop threshold corrections have
already been studied for non-decomposable ${\bf Z}_N$ Coxeter orbifolds
[\seventeen, \eightteen, \nineteen, \twentythree]. It is our purpose here
to extend the discussion to non-decomposable ${\bf Z}_M\times {\bf Z}_N$
Coxeter orbifolds. We shall find that ${\bf Z}_2\times {\bf Z}_2$ provides
the only examples.

A large class of orbifolds, the Coxeter orbifolds, can be obtained by
taking the underlying lattice of the six-torus to be a direct sum of Lie
group root lattices and constructing the generators of the point group from
Coxeter elements, generalized Coxeter elements, or their powers for the
various root lattices. A Coxeter element is the product of all the Weyl
reflections for the root lattice. When the Dynkin diagram possesses an outer
automorphism, we can also make a generalized Coxeter element by using those
Weyl reflections associated with points in the Dynkin diagram that are not
permuted by the outer automorphism together with one of the permuted Weyl
reflections and the outer automorphism itself [\twentyfour].

To construct  ${\bf Z}_M\times {\bf Z}_N$ Coxeter orbifolds, we need to
choose Coxeter elements, generalized Coxeter elements, or their powers
acting on the direct sum of root lattices, such that their eigenvalues give
the correct action of the generators of  ${\bf Z}_M$ and $ {\bf Z}_N$ in the
complex orthogonal space basis, and such that the generators of  ${\bf Z}_M$
and ${\bf Z}_N$ commute. Using only Coxeter elements,  we find no examples
for non-decomposable  ${\bf Z}_M\times {\bf Z}_N$ orbifolds. However, if we
also deploy generalized Coxeter elements, there are examples in the case of
${\bf Z}_2\times {\bf Z}_2$ point group.

The generators $\theta$ and $\omega$ of  ${\bf Z}_2\times {\bf Z}_2$ have
the action on the space basis
$$\theta=(-1, 1,-1),\qquad \omega=(1,-1,-1).\eqn\act$$
A non-decomposable example, in the sense discussed earlier, is obtained
using the lattice $SU(4)\times SO(4)\times SU(2)$. Then, in terms of Coxeter
elements, the action on the $SU(4)$, $SO(4)$ and $SU(2)$ sub-lattices is
given by $\Big({\cal C}^2(SU(4)), {\cal C}(SO(4)), 1\Big)$ and $\Big({\cal
C}^3(SU(4)
^{[2]}), I, -1\Big)$ for $\theta$ and $\omega$ respectively, where ${\cal C}$
denotes the Coxeter element, and ${\cal C}(SU(4)^{[2]})$ is the generalized
Coxeter element for the $SU(4)$ lattice. Other examples are obtained by
replacing  $SO(4)$ by $SO(5)$, $G_2$ or $SU(3)$ and  ${\cal C}(SO(4))$ by
${\cal C}^2(SO(5)), $ ${\cal C}^3 (G_2)$ or ${\cal C}^3(SU(3)^{[2]})$. However,
all of these have same action on the 2-dimensional sub-lattices as ${\cal C}
(SO(4))$ and consequently there is no relevant distinction between these
possibilities for our purposes. We shall focus on the $SO(4)$ case in what
follows.

The action of $\theta$ and $\omega$ on the basis vectors $e_1,\cdots, e_6$
of the lattice is then
$$\eqalign{\theta: &e_i\rightarrow \Theta_{ij}e_j,\quad
{\bf\Theta} \, =\, diag\Big(\pmatrix{0&0&1\cr -1&-1&-1\cr 1&0&0},
-I_2, 1\Big)\cr
\omega: &e_i\rightarrow \Omega_{ij}e_j,\quad
{\bf\Omega}\, = \, diag\Big(\pmatrix{-1&0&0\cr 0&-1&0\cr 0&0&-1},
I_2, -1\Big).}\eqn\sand$$

where $I_2 $ is the $2\times 2 $ identity matrix, and
the eigenvalues of ${\cal C}^2(SU(4))$ are $(-1,-1,1)$. It is therefore clear
that the fixed plane in the $\theta$ twisted sector lies partly in the
$SU(4)$ sub-lattice and partly in the $SU(2)$ sub-lattice so that the
${\bf T}^2\bigoplus{\bf T}^4$ decomposition does not occur.

The moduli that occur in the string loop threshold corrections to the gauge
coupling constants are those associated with fixed planes in some twisted
sectors [\four] (the ${\cal N}=2$ moduli.) In the present case, the moduli
$(T_1, U_1)$, $(T_2, U_2)$ and $(T_3, U_3)$ are associated with fixed planes
in the $\omega$, $\theta$ and $\theta\omega$ twisted sectors, respectively.
The $\omega$ twisted sector fixed plane lies in the $SO(4)$ sub-lattice,
Consequently, the $T_1$ and $U_1$ dependent threshold corrections have the
standard [\four] full modular symmetry. For the $\theta$ twisted sector,
the boundary conditions requires the winding number ${\bf w}$ and momentum
${\bf p}$ on the orbifold to satisfy\footnote*{${\bf Q}^*= {{\bf Q}^t}^{(-1)}
$}
$${\bf Q}{\bf w}\, =\, {\bf w}, \qquad {\bf Q}^*{\bf p}\, =\, {\bf
p},\eqn\sea$$
where ${\bf Q}\equiv {\bf \Theta}^t$. Thus, for this sector
$${\bf w}\, =\, \pmatrix{n_1\cr 0\cr n_1 \cr0 \cr 0 \cr n_6},\quad
{\bf p}\, =\, \pmatrix{m_1 \cr -m_1 \cr m_1 \cr 0 \cr 0 \cr m_6},\eqn\dw$$
where $m_1$, $m_6$, $n_1$ and $n_6$ are arbitrary integers.
For the $\theta\omega$ twisted sector, the boundary conditions requires
$${\bf Q}{\bar{\bf Q}}{\bf w}\, =\, {\bf w}, \qquad {\bf Q}^*{\bar{\bf
Q}^*}{\bf p}
\, =\, {\bf p},\eqn\dwe$$
where ${\bar{\bf Q}}\equiv {\bf\Omega}^t$,  and the corresponding windings
and momenta are
$${\bf w}\, =\, \pmatrix{n'_1 \cr n'_2 \cr n'_2-n'_1 \cr 0 \cr 0 \cr 0},\quad
{\bf p}\, =\, \pmatrix{m'_1 \cr m'_2 \cr -m'_1\cr 0 \cr 0 \cr 0},\eqn\sco$$
where $m'_1$, $m'_2$, $n'_1$ and $n'_2$ are arbitrary integers.

The string loop threshold corrections $\Delta_a$ may now be calculated from
partition functions for ${\cal N}=2$ twisted sectors $(h, g)$ of the orbifold,
where $h$ and $g$ refer to twists in the $\sigma$ and $t$ directions,
respectively, with both $h$ and $g$ leaving fixed the same complex plane.
It is convenient [\sixteen] to perform the calculation in terms of a subset
of ${\cal N}=2$ twisted sectors $(h_0, g_0)$, referred to as the fundamental
elements, then,
$$\Delta_a=\sum_{(h_0, g_0)}b_a^{(h_0,g_0)}
\int_{{\tilde{\cal F}}}{d^2\tau\over\tau_2}{\cal Z}_{(h_0, g_0)}(\tau,
\bar\tau)-b_a^{{\cal N}=2}
\int_{ {\cal F}}{d^2\tau\over\tau_2}, \eqn\thr$$
\noindent
where ${\cal Z}_{(h, g)}$ denotes the moduli dependent parts of the partition
functions, $b_a^{(h, g)}$ is the contribution of the $(h, g)$ sector to the
one loop renormalization group equation coefficients, $b_a^{{\cal N}=2}$ is the
contribution of all ${\cal N}=2$ twisted sectors, ${\cal F}$ is the fundamental
region for the world sheet modular group $PSL(2, Z)$, and
$\tilde{\cal F}$ is the fundamental region for the world sheet modular
symmetry group of ${\cal Z}_{(h_0, g_0)}$. Here, the single twisted sector
$(h_0, g_0)$ replaces a set of twisted sectors that can be obtained from it
by applying $SL(2, Z)$ transformations which generate
${\tilde{\cal F}}$ from ${\cal F}.$
In the present case, the fundamental elements for the non-decomposable fixed
planes are $(I, \theta)$ and $(I, \theta\omega).$
The partition functions ${\cal Z}_{(I, \theta)}$ ${\cal Z}_{(I, \theta\omega)
}$
may now be derived using methods for non-decomposable orbifolds discussed
elsewhere [\seventeen-\nineteen] with the result
$$\eqalign{&{\cal Z}_{(I, \theta)}\, =\, \cr &\sum_{\textstyle{m_1, m_6}\atop
\textstyle{n_1, n_6}} e^{\textstyle{2\pi i\tau(2m_1n_1+m_6n_6)}}\,
e^{-{\textstyle\pi\tau_2\over\textstyle Im T_2Im U_2}\,  \textstyle
\vert T_2U_2n_6+T_2n_1-2U_2m_1+m_6\vert ^2}\cr
&\cr
&{\cal Z}_{(I, \theta\omega)} \,= \, \cr
&\sum_{\textstyle{m'_1, m'_2}\atop
\textstyle{n'_1, n'_2}} e^{\textstyle{2\pi i\tau(2m'_1n'_1+\hat {m'_2}n'_2)}}
\,
e^{-{\textstyle\pi\tau_2\over\textstyle Im T_3 \, Im U_3} \textstyle\vert
T_3U_3n
'_2+T_3n'_1-2U_3m'_1+\hat{m'_2}\vert ^2},}\eqn\thr$$
where $\hat{m'_2}=m'_1-m'_2$. After Poisson resummation of the partition
functions and $\tau$ integration [\four, \seventeen, \nineteen] we find
$$\eqalign{\Delta_a \, =\, &-b_a^{(I,\theta)} log\Big(k\ Im
T_2\vert\eta(T_2)\vert^4
Im\  U_2\vert\eta(2U_2)\vert^4 \Big) \cr
-&b_a^{(I, \theta)} log\Big(k\ Im T_2\vert\eta({T_2\over2})\vert^4
Im\ U_2\vert\eta(U_2)\vert^4 \Big) \cr
-&b_a^{(I,\theta\omega)} log\Big(k\ Im T_3\vert\eta({T_3})\vert^4
Im\ U_3\vert\eta(2U_3)\vert^4 \Big) \cr
-&b_a^{(I,\theta\omega)} log\Big(k\ Im T_3\vert\eta({T_3\over2})\vert^4
Im\  U_3\vert\eta(U_3)\vert^4 \Big) \cr
-&b_a^{(I,\omega)} log\Big(k\ Im T_1\vert\eta({T_1})\vert^4
Im\  U_3\vert\eta(U_1)\vert^4 \Big) }\eqn\bos$$
where $\eta$ is the Dedekind function, and $k={\textstyle8\pi\over\textstyle
3\sqrt{3}}e^{\textstyle(1-\gamma_E)},$ where
$\gamma_E$ the Euler-Mascheroni constant.
The threshold correction $\Delta_a$ is invariant under the target space
modular symmetry group
$$\Gamma \, = \, [SL(2, Z)]_{T_1}\times [SL(2,
Z)]_{U_1}\times[\Gamma^0(2)]_{T_2}
 \times[\Gamma_0(2)] _{U_2}
\times[\Gamma^0(2)]_{T_3}\times[\Gamma_0(2)]_{T_3}
\eqn\sym$$
where $\Gamma_0(n)$ and $\Gamma^0(n)$ are congruence subgroups of $SL(2, Z)$
transformations defined by
$$T\rightarrow {aT+b\over cT+d}, \quad \hbox{with}\ c=0\ (mod\ n)\ \hbox{and}
\  b=0\ (mod\ n),\eqn\cg$$
respectively.

The string loop threshold corrections may now be used to construct the
effective potential due to  duality invariant gaugino condensates [\nine,
\sixteen]. In general, an effective potential with a realistic minimum for
the dilaton expectation value, and so realistic values for the gauge coupling
constants, is not obtained in the absence of hidden matter sector. When the
hidden sector matter is present, the requirement that the effective
superpotential should have the correct modular weight can prevent
the occurrence of such a realistic minimum when there are some ${\cal N}=1$
moduli in the theory [\fifteen, \sixteen], $i.e.,$ when there are some
complex planes which are rotated by the action of the point group in all
twisted sectors. This difficulty does not occur here because all moduli
are ${\cal N}=2$ moduli.

A simple model for the hidden sector matter [\fifteen]  is to have two
factors in the hidden sector gauge group (two gaugino condensate) with
the matter charged under these two factors coupled to singlet scalars $A_1$
and $A_2$ with self couplings $A_1^3$ and $A_2^3$. In such a model,
the perturbative superpotential $W_p$ is given by
$$W_p \, =\, MA_1\sum_{\alpha}Q_\alpha{\bar
Q}_{\alpha}+MA_2\sum_{\beta}R_{\beta}
{\bar R}_\beta+A_1^3+A_2^3\eqn\shahadi$$
where $Q_\alpha$ and $R_\beta$ are the matter fields coupled to the two
factors of the gauge group, with $M$ is the string scale.
The renormalization group equations for the hidden sector gauge coupling
constants $g_a(\mu)$, $a=1, 2$, at scale $\mu$, are
$$g_a^{-2}(\mu)\, = \, g_a^{-2}(M)+ {{(b_a)}_0\over16\pi^2}\, log
{{(M_a)}^2_I\over
\mu^2}+{{(b_a)}\over16\pi^2}\, log {{(M)}^2\over {(M_a)}_I^2}+{\Delta_a
\over16\pi^2}\eqn\steve$$

where ${(b_a)}_0$ is the renormalization group coefficient for the pure
gauge case, ${b_a}$ is the complete renormalization group coefficient,
${(M_a)}_I$ are the intermediate scales
$${(M_a)}_I \, =\, M <A_a >, \qquad a=1, 2\eqn\sop$$
In general, when the modular group symmetries are congruence subgroups of
$SL(2, Z)$ the string loop threshold corrections have the form [\seventeen,
\nineteen]
$$\eqalign{\Delta_a \, =\,
&-\sum_i({{b'}_a}^i-\delta_{GS}^i)\Big(log({{T_i+{\bar T}_i}\over l_{im}}
)+\sum_m{c_{im}\over2}log\vert{\eta(T_i)\over l_{im}}\vert^4\Big)
\cr &-\sum_i(\tilde{{b'}}_a^i-{\tilde\delta}_{GS}^i)\Big(log({{U_i+{\bar U}_i}
\over {\tilde l}_{in}} )
+\sum_n{{\tilde c}_{in}\over2} log\vert{\eta(U_i)\over {\tilde l}_{in}}\vert
^4\Big)}\eqn\alex$$
where ${b'_a}^i$, $\delta_{GS}^i$, ${\tilde{b'}}_a^i$ and ${\tilde\delta}_
{GS}^i$
are the usual [\six, \seven ] duality anomaly and Green-Schwarz coefficients
for the $T$ and $U$ moduli, $l_{im}$ and ${\tilde l}_{in}$ are rational
numbers, and the integer-valued coefficients
$\tilde{c_{in}}$ and ${c_{im}}$ satisfy
$$\sum_n \tilde{c_{in}}=\sum_m {c_{im}}=2.\eqn\hiam$$
(This ensures that the corresponding non-perturbative superpotential has
the correct modular weights). In \alex, and what follows we have replaced
the original $T_i$ and $U_i$ by $-iT_i$ and $-iU_i$ to conform with the
notations usually employed in the discussion of effective potentials.

The gauge kinetic function may be read off from \steve\ and \alex, and
the corresponding two-gaugino condensate non-perturbative superpotential
$W_{np}$ takes the form
$$W_{np}=\sum_{a=1,2}d_a e^{{\textstyle 24\pi^2}{\textstyle S}
\over{\textstyle{(b_a)}_0}}A_a^
{\mu_a}H_a(T_i, U_i),\eqn\royal$$
where $S$ is the dilaton field,
$$\mu_a={3\Big({(b_a)}_0-b_a\Big)\over {(b_a)}_0}, \quad d_a={b_a\over
96\pi^2g_a}\eqn\delft$$
and
$$H_a(T_i, U_i)=\prod_{i,m}{\textstyle\Big(\eta({\textstyle T_i\over
\textstyle l_{im}})
\Big)^{ - {\textstyle
3c_{im}({b'_a}^i-\delta_{GS}^i)\over{\textstyle(b_a)}_0}}}
\prod_{i,m}\Big(\eta({\textstyle U_i\over\textstyle {\tilde l}_{in}})
\Big)^{-{\textstyle 3{\tilde c}_{in}({\tilde{b'}_a}^i-{\tilde\delta}_{GS}^i)
\over \textstyle{(b_a)}_0}}\eqn\christine $$
The complete superpotential is
$$W=W_{np}+W_p,\eqn\bei$$
where $$\delta_{GS}^i={\tilde\delta}_{GS}^i=0, \quad \ \hbox{for\ all\ } i
\eqn\need$$
as is true for the ${\bf Z}_2\times {\bf Z}_2$ orbifold. After `integrating
out'
the singlet scalars $A_1$ and $A_2$,
we obtain for the effective potential
$$\eqalign{&(S+\bar S)\prod_i(T_i+\bar T_i)(U_i+\bar U_i)V_{eff}=
\vert W_{eff}-(S+\bar S){\partial W_{eff}\over \partial S}\vert^2+\cr
&\sum_i\Big\{\vert W_{eff}-(T_i+\bar T_i){\partial W_{eff}\over \partial T_i}
\vert^2+\vert W_{eff}-(U_i+\bar U_i){\partial W_{eff}\over \partial U_i}
\vert^2\Big\}-3\vert W_{eff}\vert^2,}\eqn\paul$$
with $$W_{eff}=\Omega(S)\Psi(T_i,U_i)\eqn\hell$$
where
$$\eqalign{&\Omega(S)=-\sum_a {b_a\over\Big(b_a-{(b_a)}_0\Big)}
\Big(-{\mu_ad_a\over3}\Big)^{\textstyle{(b_a)}_0\over\textstyle b_a}
e^{\textstyle{24\pi^2\over\textstyle b_a}S},\cr
&\Psi(T_i,U_i)=\prod_{i,m}\Big(\eta({T_i\over l_{im}})
\Big)^{-\textstyle c_{im}}\prod_{i,n}\Big(\eta({U_i\over {\tilde l}_{in}})
\Big)^{-\textstyle{\tilde c}_{in}}}\eqn\sidon$$
In the present case, \bos\ implies that
$$\Psi(T_i,U_i)=\Big(\eta(T_1)\eta(U_1)\Big)^{-2}
\prod_{j=2,3}\Big(\eta(T_j)\eta({T_j\over2})\eta(U_j)\eta(2U_j)\Big)^{-1}\eqn
\pinhead$$

Equation \paul\ has the usual extremum for $S$ at
$$\Omega(S)-(S+\bar S){d\Omega\over dS}=0,\eqn\pain$$
and it is known [\fifteen] that this can give a realistic minimum for $S$
with $S_R\approx 2$ for many possible $SU(N_1)\times SU(N_2)$ hidden sector
gauge group with hidden sector matter in fundamental $SU(N_1)$ and $SU(N_2)$
representations and their conjugates. At the minimum for $S$,
$$\eqalign{&\vert\Omega(S)\vert^{-2}(S+\bar S)\prod_i(T_i+\bar T_i)
(U_i+\bar U_i)V_{eff}=\cr
&\sum_i \Big\{ \vert \Psi-(T_i+\bar T_i){\partial \Psi\over \partial
T_i}\vert^2 +
\vert  \Psi -
(U_i+\bar U_i){\partial \Psi\over \partial U_i}
\vert^2\Big\}-3\vert\Psi\vert^2}\eqn\bas$$
Because $\delta_{GS}^i$ and  ${\tilde \delta}_{GS}^i$ are zero, there are
extrema for $T_i$ and $U_i$ at the self-dual (or fixed) points of the modular
symmetry
group \sym,
$$T_1=1\ \hbox{or}\  e^{i\pi\over6}, \quad
U_1=1\ \hbox{or}\ e^{i\pi\over6}, \quad T_2=T_3=1-i, \quad U_2=U_3={(1+i)
\over2}.\eqn\susy$$
These extrema  are saddle points and maxima for the fixed points in $(T_1, U_1
) $,
and maxima at the remaining fixed points listed in eqn. \susy .

A numerical minimization of $V_{eff} $ shows that the minima for $T_i$ and
$U_i$ occur at
$$\eqalign{T_1&= U_1= 1.26, \quad T_2= 1.41 , \cr
 U_2 & =0.71 + i , \quad T_3= 1.41 , \quad U_3=0.71 +  i.}
\eqn\min$$
This provides an elegant example of an orbifold for which there is an
anisotropic solution for the moduli.

To date, it is not known how to calculate explicit string loop threshold
corrections in the presence of Wilson lines. However, it is possible to
identify modular symmetries that those threshold corrections will possess,
and a set of conditions to determine these modular symmetries has been
written down elsewhere [\twentytwo, \twentythree]. For the ${\bf Z}_2\times
{\bf Z}_2$ case, the inequivalent Wilson lines ${\tilde a}_{bI}$, $b=1,\cdots,
6$, $I=1,\cdots,16$, satisfy
$${\tilde a}_{3I}={\tilde a}_{1I}, \qquad 2{\tilde a}_{bI}\in \Lambda_
{E_8\times E'_8}\eqn\wils$$
The matrices $A^t$ of refs [\twentytwo, \twentythree] can be constructed
from
${\tilde a}_{bI}$,
$$A^t_{bB}={\tilde a}_{bI}E^I_B\eqn\bA$$
where $E^I_B$ is a vielbein for the $E_8$ lattice, and the matrices $K_1$,
$K_2$ and $K_3^t$ of ref [\twentythree] are given by
$$K_1=\pmatrix{2&0\cr 0&-1},K_2=\pmatrix{
1&0&0&0&0&0\cr
0 &0&0&0&0&0 \cr
1 &0&0&0&0&0 \cr
0 &0&0&0&0&0\cr
0 &0&0&0&0&0\cr
0 &0&0&0&0&1},
K^t_3=\pmatrix{1&0&0&0&0&0\cr
0 &0&0&0&0&1 \cr
0 &0&0&0&0&0 \cr
0 &0&0&0&0&0\cr
0 &0&0&0&0&0\cr
0 &0&0&0&0&0},\eqn\spiel$$
 in the $\theta$ sector
and $$K_1=\pmatrix{2&0\cr 1&-1},
K_2=\pmatrix{
1 &0&0&0&0&0\cr
0 &1&0&0&0&0 \cr
-1 &1&0&0&0&0 \cr
0 &0&0&0&0&0\cr
0 &0&0&0&0&0\cr
0 &0&0&0&0&0},
K^t_3=\pmatrix{
1 &0&0&0&0&0\cr
0 &1&0&0&0&0 \cr
0 &0&0&0&0&0 \cr
0 &0&0&0&0&0\cr
0 &0&0&0&0&0\cr
0 &0&0&0&0&0}\eqn\nest$$
in the $\theta\omega$ sector.
The modular symmetry groups for $T_1$ and $U_1$ may now be deduced
from (48)-(55) of ref [\twentytwo], or (5.15) and (5.16) of ref
[\twentythree]. The modular symmetry groups for $T_2$, $U_2$, $T_3$ and
$U_3$ may be deduced from (6.13) and (6.14) of ref [\twentythree]. A simple
example is obtained by taking
$$A^t= {1\over2}\left (
\matrix{1&1&0&0&0&0&0&0\cr
        1&1&0&0&0&0&0&0\cr
        1&1&0&0&0&0&0&0\cr
        0&0&0&0&0&0&0&0\cr
        0&0&0&0&0&0&0&0\cr
        0&0&0&0&0&0&0&0 }\right )\eqn\examp$$
in which case the modular symmetry groups associated with the various $T$ and
$U$ moduli
are as follows. Define the  subgroup $ \tilde{\Gamma}_0(2) $
 of the congruence subgroup
$\Gamma_0(2) $ as in eqn.\cg , but with the additional constraint
$b\, =\, 0\,  (mod 2) $. Then for $T_1\, , \, U_1 $ we obtain
$[PSL(2,Z)]_{T_1}\, ,
\, [PSL(2,Z)]_{U_1}$,
for $T_2 \, , \,  U_2 $, $ [\tilde{\Gamma}_0(2)]_{T_2}\, , \,
[\tilde{\Gamma}_0(2)]_{U_2} $ and finally for $T_3 \, , \,  U_3 $ we find the
subgroups $ [\tilde{\Gamma}_0(2)]_{T_3}\, , \,  [\Gamma_0(4)]_{U_3} $

In conclusion, we have shown that ${\bf Z}_2\times {\bf Z}_2$ is the only
 ${\bf Z}_M\times {\bf Z}_N$  Coxeter orbifold with a choice of lattice for
 which the modular symmetry groups, in the absence of Wilson lines, are
 congruence subgroups of $SL(2, Z)$ rather than the full modular group. The
 explicit string loop threshold corrections displaying these modular
 symmetries have been calculated and used to derive the values of the $T$
 and $U$ moduli by minimizing duality invariant effective potential. Finally,
 we have discussed possible modular symmetries of threshold corrections for
 ${\bf Z}_2\times {\bf Z}_2$ orbifolds with Wilson lines background.

\centerline{\bf{ACKNOWLEDGEMENT}}

This work is supported in part by P.P.A.R.C. and the work of S.T.  is
supported by the Royal Society.

\refout
\end